# ECG-Based Driver Stress Levels Detection System Using Hyperparameter Optimization

**Mohammad Naim Rastgoo[1,2], Bahareh Nakisa[3], Andry Rakotonirainy[2], Frederic Maire[1], Vinod Chandran[1]**

[1] School of Electrical Engineering and Robotics, Queensland University of Technology, Brisbane, QLD, Australia
[2] Centre for Accident Research & Road Safety Queensland, Queensland University of Technology, Brisbane, Qld, Australia
[3] School of Information Technology, Deakin University, Waurn Ponds, VIC, Australia

Corresponding author: Mohammad Naim Rastgoo (e-mail: Mohammadnaim.rastgoo@qut.edu.au).

**ABSTRACT** Stress and driving are a dangerous combination which can lead to crashes, as evidenced by the large number of road traffic crashes that involve stress. Motivated by the need to address the significant costs of driver stress, it is essential to build a practical system that can classify driver stress level with high accuracy. However, the performance of an accurate driving stress levels classification system depends on hyperparameter optimization choices such as data segmentation (windowing hyperparameters). The configuration setting of hyperparameters, which has an enormous impact on the system performance, are typically hand-tuned while evaluating the algorithm. This tuning process is time consuming and often depends on personal experience. There are also no generic optimal values for hyperparameters values. In this work, we propose a meta-heuristic approach to support automated hyperparameter optimization and provide a real-time driver stress detection system. This is the first systematic study of optimizing windowing hyperparameters based on Electrocardiogram (ECG) signal in the domain of driving safety. Our approach is to propose a framework based on Particle Swarm Optimization algorithm (PSO) to select an optimal/near optimal windowing hyperparameters values. The performance of the proposed framework is evaluated on two datasets: a public dataset (DRIVEDB dataset) and our collected dataset using an advanced simulator. DRIVEDB dataset was collected in a real time driving scenario, and our dataset was collected using an advanced driving simulator in the control environment. We demonstrate that optimising the windowing hyperparameters yields significant improvement in terms of accuracy. The most accurate built model applied to the public dataset and our dataset, based on the selected windowing hyperparameters, achieved 92.12% and 77.78% accuracy, respectively.

## I. INTRODUCTION

Driving is a perceptual motor skill that involves multiple situations and consequently results in varying levels of stress. Stress can increase crash risk nearly tenfold according to Virginia Tech Transportation Institute (Brown et al., 2016). Australian national crash reports also show that feeling stressed is a contributing factor to fatal crashes (Beanland, Fitzharris, Young, & Lenné, 2013). Thus, stress is an important factor in driving which can result in poor driving performance and reduce road safety. A key strategy to enhance safety is to provide an in-vehicle assistance system that can detect a driver's stress level with high accuracy.

It is generally accepted that analyzing the autonomic nervous system (ANS) activity is essential for building a stress level detection system. ANS activities related to stress are categorized into momentary and dynamic activities which are monitored through physiological body responses. Most of the studies in the domain of driver stress classification use several physiological signals to monitor drivers' stress levels (J.A. Healey & Picard, 2005; Soman, Sathiya, & Suganthi, 2014; Lanatà et al., 2015). Despite the fact that a fusion of physiological signals has been shown to be an efficient and effective approach to build a reliable driver stress detection system (J. Healey & Picard, 2000; Katsis, Katertsidis, Ganiatsas, Fotiadis, & others, 2008; Singh, Conjeti, &

Banerjee, 2013), this approach presents a serious challenge to acceptability of the system by drivers in real-world situations. This is because wearing several body contact physiological sensors is not only obtrusive, but also can easily distract the driver and decrease road safety (Rastgoo et al., 2018). Regarding the acceptability issue, building a driver stress level detection system based on a single physiological modality has received much attention in recent years. Electrocardiogram (ECG) signal, which represent the heart electrical activity over time, is known as a reliable and accurate physiological indicator for driver stress classification (Rastgoo et al., 2018). Heart rate variability (HRV) that is directly extracted from ECG signal, is an important heart activity parameter related to stress.

Although ECG signal is a reliable stress indicator, extracting high quality features to build an accurate model to classify driver's stress levels is a challenging task. The process of building driver stress classification model using ECG signal is divided into four main steps: pre-processing, signal segmentation/ windowing, feature extraction, and classification. In the first step, the raw ECG signal is pre-processed using noise reduction techniques, and then low-level features such as HRV parameters are extracted from the cleaned ECG data. In the second step (signal segmentation/ windowing), the cleaned raw ECG signal or low-level extracted features are divided into sequences of windows with a fixed size and a certain degree of overlap between the adjacent windows. Then, a set of higher-level features are extracted from each window and finally fed into a classifier to discriminate different stress levels.

Most of the studies in the domain of stress detection focused on extracting new features or applying the state-of-the-art classifiers and used the traditional method (manual hyperparameter selection) for signal segmentation (second step) (Wang, Lin, & Yang, 2013). In the manual hyperparameter optimization approach, the hyperparameter values are selected by an expert. However, this approach can potentially lead to the selection of non-optimal hyperparameters values and reduce the accuracy of classification models. In the process of extracting features, the signal segmentation/ windowing step plays a key role. This is because the quality of the extracted features and consequently the performance of the classifier depend on the windowing hyperparameter values (window size and degree of overlap values). Moreover, there are no generic optimal hyperparameter values that can be used for different classification problems. It is therefore important to identify the best possible values for these hyperparameters to achieve an accurate ECG-based driver stress level detection model.

Another approach is automatic hyperparameter selection, which employs optimization techniques to find appropriate hyperparameters values. Although optimization techniques suffer from high computational costs, the techniques guarantee near optimal solutions which can lead to higher classification performance. Therefore, an automatic hyperparameter approach is a preferred choice to obtain appropriate values and increase the performance of detection system.

One of the efficient optimization algorithms is particle swarm optimization (PSO). The PSO algorithm can optimize the hyperparameters by assessing different values. The key advantages of using the PSO algorithm to solve optimization problem is that it is easy to implement, highly efficient in its search strategy and can convergence fast (Lorenzo et al., 2017a; Bahareh Nakisa, Ahmad Nazri, et al., 2014; Bahareh Nakisa, Rastgoo, et al., 2014a). In this work, we propose a framework based on the PSO algorithm to optimize the windowing hyperparameters (window size and the degree of overlap) and improve the performance of ECG-based driver stress level detection models. To the best of our knowledge, this is the first systematic study to build an accurate ECG-based stress detection model based on optimising windowing hyperparameters. The proposed method is evaluated on two different datasets, a public dataset (DRIVEDB dataset) and our dataset. DRIVEDB dataset was collected in a real time driving scenario (J. Healey & Picard, 2000), and our dataset was collected using an advanced driving simulator. Moreover, to evaluate the efficiency of the proposed framework, different ECG feature sets from time and frequency domains are considered. Evaluating the performance of the proposed framework on different feature sets can help to identify suitable hyperparameter values for each feature set. The following are the primary contributions of this study:

- Proposing an enhanced framework for driver stress level classification that includes the optimization of windowing hyperparameter values. This study shows that optimising windowing hyperparameters (window size and the degree of overlap) using the PSO algorithm can significantly improve the performance of driver's stress level classification.

- Evaluating the performance of the proposed framework using state-of-the-art hyperparameter optimization methods (Random search and PSO) on two datasets. We use a public dataset, collected in real time driving scenarios, and our dataset, collected using an advanced driving simulator. The effectiveness of the proposed framework is also evaluated using different ECG feature sets to identify the suitable windowing hyperparameters values for each feature set.

- Conducting various experiments to compare the performance of a driver stress classification system designed using the proposed framework with other existing techniques on two datasets. We show that the proposed framework results in better performance than previously published results for three stress level classification of driver stress.

This paper is structured as follows: section II presents an overview of related work. Section III discusses the proposed framework. Section IV presents the experimental method, including the data collection and adjustment of the algorithm. Section V presents and discusses the experiment results. Finally, section VI presents our conclusions and suggestions for future work.

## II. RELATED WORK

One of the crucial steps in the process of building an accurate driver stress level detection model is signal segmentation/windowing. Segmentation corresponds to the process of dividing signals into smaller segments and these smaller segments are called window sizes. Several studies take advantage of combining physiological signals using different window sizes to build an accurate driver stress detection model (Healey & Picard, 2000; Katsis et al., 2008; R. R. Singh et al., 2013; Urbano, Alam, Ferreira, Fonseca, & Simíões, 2017). However, none of the studies consider the degree of overlap between successive windows. Urbano et al. (2017) fused nine features extracted from ECG and electrodermal activity (EDA) signals to detect two stress levels. The extracted features were fed to a linear discriminant analysis classifier to build a personal model for six drivers. The built models achieved 81–97% accuracy. In the study, the signals were segmented into 60-second windows with no degree of overlap. In another study (R. R. Singh et al., 2013), photoplethysmogram (PPG), EDA and respiration (RSP) signals were segmented into 10-second windows with no degree of overlap. The extracted features from the signals were fed to a recurrent neural network classifier to discriminate three levels of drivers' stress. For a group of 19 drivers, the precision, sensitivity, and specificity of the proposed model were reported 89.23%, 88.83% and 94.92%, respectively. Healey and Picard (2000) collected ECG, EDA and RSP signals from 10 drivers to detect four levels of stress using a k-nearest neighbor (KNN) classifier. The collected signals were segmented into 60-second windows with no degree of overlap. The built model achieved 86% accuracy. In another study (Katsis et al., 2008),

12 features were extracted from EDA, ECG, RSP and electromyography (EMG) signals from 10 drivers to detect four stress levels using a support vector machine (SVM) classifier. The signals were segmented into 10-second windows with no degree of overlap. The proposed model achieved 79.3% detection accuracy.

Although building multimodal model based on the fusion of different physiological signals is successful to classify driver's stress levels, there are some serious challenges for acceptability and usefulness of this system for real-world applications. This is because most of the physiological signals are recorded by body contact sensors. Since wearing several body contact sensors simultaneously restricts drivers' movements, increases their awareness of being monitored and easily distracts them, the system using multiple modalities may not be practical for continuous driver stress monitoring. Therefore, to build an accurate driver stress level classification model with high system acceptability, it is essential to use a single physiological signal.

### A. ECG-based driver stress level detection

Among studies on physiological signals, some of them have evaluated the effectiveness of ECG signal to classify driver's stress levels (Bichindaritz et al., 2017; Keshan et al., 2015; Rastgoo, 2019; Rastgoo et al., 2019; Wang et al., 2013). Wang et al. (2013) proposed a model based on several time-domain features extracted from ECG signal to detect two stress levels of drivers. The signal was divided into 5-minute windows with 50% overlap and 56 features were extracted from each window. The proposed model is evaluated on DRIVEDB dataset and achieved 97.78% accuracy using a KNN classifier. In another study (Keshan et al., 2015), a group of statistical features were extracted from ECG signal to classify driver's stress into two and three levels. The ECG signal was segmented into three different window sizes ranging from 14 to 38 minutes, and 14 features were extracted from each window. The performance of the study using 10 classifiers was compared on DRIVEDB dataset and the highest accuracies to detect two and three stress levels were 97.92% (using a decision tree classifier), and 67.16% (using a multilayer perceptron (MLP) classifier) respectively. Bichindaritz et al. (2017) used ECG signal to detect three stress levels of drivers. The ECG signal was divided into three different window sizes, ranging from 14 to 38 minutes, and 10 features (6 statistical and 4 non-linear) were extracted from each window. The extracted features were fed into 12 classifiers. The result showed that MLP has the best performance (80.60% accuracy) among all the applied classifiers.

Based on the reviewed literature, most of the works have selected manually the window size for ECG signal. However, selection of optimal hyperparameters requires a high level of domain knowledge and expertise. In addition, the process of searching for optimal values is time consuming and error prone.

### B. Hyperparameter Optimization

As discussed earlier, most studies in the domain of stress detection used the manual hyperparameter approach to select the window size and the degree of overlap. In the manual hyperparameter approach, the hyperparameters values are selected by the expert. However, this can result in selecting non-optimal hyperparameter values and low classification performance. Another efficient and effective approach that windowing hyperparameters can benefit from is automatic hyperparameter optimization. Automatic windowing hyperparameter selection can improve the performance of driver stress level classification system. Hyperparameter optimization can be interpreted as an optimization problem where the objective is to find hyperparameter values that maximizes the performance and yield of an accurate model. The automatic hyperparameter selection approach has been successfully used in different research domains (Bergstra & Bengio, 2012; Bergstra et al., 2011).

Optimization techniques such as simple grid search, random search (J. S. Bergstra et al., 2011b; Krstajic et al., 2014) and evolutionary computation (EC) algorithms (Hutter et al., 2011; B. Nakisa et al., 2018; Qin et al., 2017) have been applied to different hyperparameters optimization problems. The grid search algorithm is one of the traditional optimization algorithms which is based on an exhaustive search. In grid search the set of trials is formed by assembling every possible combination of values. However, this algorithm takes a long period of time to find the global optimum because it searches all the possible solutions. Using grid search algorithm with large search space is not promising. Another hyperparameter selection algorithm is random search which is based on the direct search algorithm. It is a popular technique because of good prediction in low dimensional numerical input spaces. Random search first initializes random solutions and then computes the performance of the initialized random solutions. Finally, it selects the best possible solutions based on the problem objective (maximization or minimization). Random search is more efficient than grid search. Although this algorithm is easy to implement, it suffers from slow convergence (Van Der Maas et al., 2005). The main drawback of random search is that it doesn't utilize the updated information from each new trial point, and they rely on the predetermined search strategy. Random search are slightly less distributed in the original space, but far more evenly distributed in the subspaces (J. Bergstra & Bengio, 2012). Therefore, Random-based search techniques are less efficient than more sophisticated optimization methods.

Other existing approaches to optimize hyperparameters are Evolutionary Computation (EC) algorithms. These algorithms are beneficial because they are conceptually simple and can often achieve highly competitive performance in different domains (Dasgupta & Michalewicz, 2013; B. Nakisa et al., 2018; Nakisa et al., 2014; B. Nakisa et al., 2014, 2017, 2018; B. Nakisa, Rastgoo, et al., 2014b; Rastgoo et al., 2015). One of the most powerful EC algorithms is the PSO algorithm. The PSO algorithm was inspired by the social behavior of fish schooling and bird flocking (Kennedy & Eberhart, 1995). Because of its simplicity and effectiveness, PSO has been applied in different optimizations domains such as robotics (Couceiro et al., 2011; Nakisa et al., 2014a; Nakisa et al., 2014b; Rastgoo et al., 2015), job scheduling (Sha & Hsu, 2006; Zhang et al., 2009) and feature selection (Nakisa et al., 2017). Recently, PSO has been applied successfully in optimising hyperparameters such as number of inputs, hidden nodes and learning rate on classification problems (Lorenzo et al., 2017b; B. Nakisa et al., 2018; B. Nakisa, 2019; Ye, 2017).

In this study, we investigate the performance of PSO algorithm on ECG-based driver's stress levels classification and find the optimal/near optimal windowing hyperparameter values. To our knowledge, there is no study that focuses on selecting windowing hyperparameter values automatically in the domain of stress detection. This study proposes an enhanced framework based on the PSO algorithm to optimize windowing hyperparameter values (the window size and overlap hyperparameters) and improve the performance of ECG-based driver's stress level detection model.

### III. METHODOLOGY

In this section, the proposed driver stress classification framework using the PSO algorithm to optimize the windowing hyperparameters (window size and the degree of overlap) is presented.

The framework contains three main stages: pre-processing, segmentation/windowing hyperparameter optimization and driver stress detection model (see Figure 1).

In the *first stage*, the *pre-processing stage*, ECG data are cleaned, and then heart rate variability (HRV) is measured

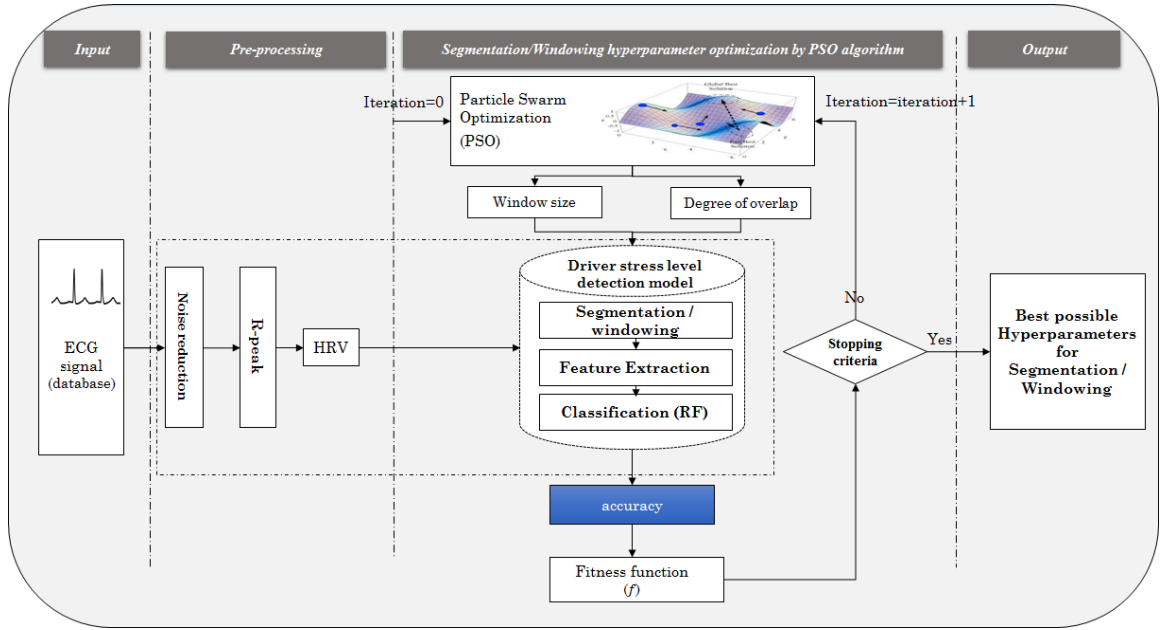

**Figure 1.** Driver stress levels classification framework using PSO-based windowing hyperparameters optimization (window size and degree of overlap between windows).

. In the *second stage*, the *PSO algorithm* is applied to find the best possible windowing hyperparameter values (window size and degree of overlap).

To evaluate the performance of the selected hyperparameter values, *a driver stress detection model* is built (*third stage*). The selected hyperparameter values are fed into the driver stress detection model to evaluate the performance of the built model based on the selected values.

The PSO algorithm explores different hyperparameter values, in an iterative process, to find the best possible values (solution). At each iteration, based on the selected hyperparameter values, an ECG-based stress level detection model, which contains feature extraction and classification steps, is built. Then, the performance of the built model, in terms of accuracy, is evaluated to determine the efficiency of the hyperparameters.

The next sections describe each stage in more detail.

### A. Pre-Processing

The ECG signal is usually weak and feature undesired noise resulting from baseline wander, muscle movement, and electrode misplacement. Therefore, in the pre-processing stage, the raw ECG data is first downsampled to 200 Hz and then a Butterworth band-pass filter (5–15 Hz) is applied to reduce the noise.

After noise reduction step, HRV data is extracted from the raw ECG signals. This due to the fact that, the relevant ECG features for driver's stress detection are generally extracted from HRV data (Rastgoo et al., 2018). HRV is defined as the time fluctuations between sequences of successive heart beats. To measure these parameters in this study, first R-peaks are extracted from the ECG signal using the Pan-Tompkins algorithm (Pan & Tompkins, 1985), and then the HRV data based on the extracted peaks are measured.

### B. Segmentation/ windowing hyperparameter optimization

After cleaning the data and extracting HRV data from the raw ECG signal, the segmentation/ windowing hyperparameter optimization stage is applied. In this process, the PSO algorithm is used to find the best possible values for the windowing hyperparameters (the window size and the degree of overlap). Based on the Figure 1, PSO algorithm starts to select the hyperparameter values randomly and passes these values to the driver stress detection model to evaluate. To assess the efficacy of the selected hyperparameter values, the accuracy of the built model is measured.

This process is done iteratively until the PSO algorithm finds the best possible values. The next step, the driver's stress level detection model, contains feature extraction and classification.

In the following subsections, first the process of adapting the PSO algorithm to find the potential windowing hyperparameter values is explained, and then the process of modelling for drivers' stress level detection is presented.

### Particle swarm optimization (PSO)

In this study, the PSO algorithm is used to find the best possible set of windowing hyperparameter values to enhance the performance of an ECG-based drivers' stress level classification system. The PSO algorithms consists of a set of particles and a fitness function.

To apply PSO algorithm for windowing hyperparameter optimization, three main steps should be followed: *initialization, evaluation, updating.*

In the first step, *initialization*, a population of particles is initialized: $x_i^t = \{x_{1,i}^t, x_{2,i}^t, \dots, x_{D,i}^t\}, i = 1, \dots, N$, where N is the population size, $t$ is the current iteration and $D$ is the dimension of problem space. In this study, PSO algorithm consists of 5 particles. The initialization iteration is zero.

Each particle starts to search with randomized positions $x_{D,i}^t$, and velocity $v_{D,i}^t$, in the $D$-dimensional problem space. As the aim of this study is to optimize the window size and the degree of overlap hyperparameters, the dimension of the problem space is two (D=2). It should be noted that each particle represents a solution. Particles are spread over the problem space randomly; therefore, particles' positions represent the window size and the degree of overlap values in the problem space. The search space is defined as between the minimum and maximum values for the window size and the degree of overlap: $x_D^{lower} \leq x_D \leq x_D^{upper}, D=1, 2$.

As we evaluate the performance of the proposed method on two datasets, DRIVEDB dataset and our dataset, we need to adapt the window size and the degree of overlap on each dataset individually. This is because the experiment duration for each dataset is different. For DRIVEDB dataset, the window size values are selected to range from 5 to 520 seconds $x_1^{lower} = 5$ and $x_1^{upper} = 520$ and the values for the degree of overlap were selected to range from 0 to 95 percent: $x_2^{lower} = 0$, $x_2^{lower} = 95$. The total data points in the problem space for DRIVEDB dataset is 48,925.

For our dataset, the window size values are selected to range from 5 to 60 seconds: $x_1^{lower} = 5$ and $x_1^{upper} = 60$, and the values for the degree of overlap were selected to range from 0 to 95 per cent: $x_2^{lower} = 0$ and $x_2^{lower} = 95$. The total data point in the problem space for our dataset is 5,225.

Once the particles are distributed into the search space, the current position of each particle (window size and the degree of overlap) needs to be evaluated. In the second step, *evaluation*, based on the current position of each particle, a stress level detection model is built to evaluate the performance (accuracy) of the built model.

In the third step, *updating*, the next position and velocity of each particle are updated using equations 1 and 2. The velocity vector (eq. 1) is calculated using the particle's personal best position ($p_{best}$), the global best position ($g_{best}$) and the current velocity vector. The particle's personal best value is the best position that each particle has visited so far and the global best ($g_{best}$) is the best visited position among all particles in the population. These two values ($p_{best}, g_{best}$) can be controlled by some learning factors. The following equations update the particle's velocity and position.

$$v_i^{t+1} = \omega v_i^t + c_1 r_1(p_{best}^t - x_i^t) + c_2 r_2(g_{best}^t - x_i^t) \qquad (1)$$

$$x_i^{t+1} = x_i^t + v_i^{t+1} \qquad (2)$$

Where $t$ represents the current search iteration, $v_i^t$ and $x_i^t$ are the current particle's velocity and position vectors respectively, $\omega$ is the inertia weight, $c_1, c_2$ are learning factors and $r_1, r_2$ are random numbers uniformly distributed between [0, 1].

The last two steps, evaluation and updating, are iteratively done until the termination criteria are met. The termination criteria could be the maximum number of iterations (30 iterations) or finding the optimal set of window size and degree of overlap that maximize the accuracy (100% accuracy) of driver stress model. Figure 2 presents the pseudo-code for windowing hyperparameter optimization using PSO algorithm.

Pseudo-code for windowing hyperparameter optimization using PSO algorithm

Define the size of the population, *N*, D dimension of problem, the inertia weight ($\omega$), learning factors ($c_1, c_2$) and random numbers ($r_1, r_2$)

**Initialization**: Initialize the population $x_i^t = \{x_{1,i}^t, x_{2,i}^t, \dots, x_{D,i}^t\}, i = 1, \dots, N$. Each particle is uniformly distributed in the range of $x_D^{lower} \leq x_D \leq x_D^{upper}, D = 1, 2$.

*While* the stopping criteria are not met
    *For* each individual particle, in the population N

        ***Evaluation:*** The current position of the particle should be evaluated. Based on the current position of each particle, a Drivers' stress level detection model is built.
        *Driver stress detection model* consists of Segmentation, feature extraction and classification steps.

        ***Updating:*** The next position and velocity of each particle are computed using the following equations:

        $v_i^{t+1} = \omega v_i^t + c_1 r_1(P_{best}^t - x_i^t) + c_2 r_2(g_{best}^t - x_i^t)$

        $x_i^{t+1} = x_i^t + v_i^{t+1}$
    *End For*
  *End While*

**Figure2.** The pseudo-code for windowing hyperparameter optimization using PSO algorithm.

**Table 1.** The proposed ECG feature sets

| Feature set | | Domain | Extracted features |
|---|---|---|---|
| Statistical HRV features | | Time | Mean of RR, standard deviation RR, mean absolute of RR, SDNN, RMSSD, NN20, NN50, and PNN50 |
| Non-linear HRV features | | Time | SD1, SD2, and SD1/SD2 |
| Frequency-domain HRV features | | Frequency | HF, LF, and LF/HF |

### Drivers' stress level detection model

To evaluate the performance of generated solutions (window size and the degree of overlap) using the PSO algorithm, a drivers' stress level detection model is built. The process of building the model is explained in more detail in the next section.

### Segmentation and feature extraction

The ECG low-level data (HRV feature) is segmented based on the selected window size and the degree of overlap. Then, high-level ECG features are extracted. In our previous work (Rastgoo et al., 2018), a list of ECG features extracted from HRV data in relation to drivers' stress levels is reviewed. To optimize the feature extraction process, the most widely used ECG features are extracted (tabulated in Table 1). These features are extracted from the time and frequency domains. In the *HRV time domain features,* there is a significant negative correlation between the HRV signal and the driver stress level (Eilebrecht et al., 2012; Miller & Boyle, 2013). The most common extracted time-domain features from the HRV data to use for drivers' stress detection are the mean of first difference, average normal-to-normal (NN) and intervals, standard deviation of normal-to-normal intervals (SDNN), square root of the mean squared difference of successive normal-to-normal intervals (RMSSD), and number of pairs of successive normal-to-normal intervals that differ by more than 50 MS (PNN50) (J. Healey & Picard, 2000; J.A. Healey & Picard, 2005; Katsis et al., 2008; Lanatà et al., 2015; Munla et al., 2015; Wang et al., 2013). Poincaré plot is one of the most important non-linear parameters of the HRV signals. Two important statistical Poincaré plot features in relation to stress level detection are the standard deviation of the short-term HRV, known as SD1, and the standard deviation of the HRV, known as SD2. In the *HRV frequency-domain features*, the most widely used parameters to detect drivers' stress levels are high frequency (HF) power spectrum (ranging from 0.2 to 0.4 Hz) and low frequency (LF) power spectrum (ranging from 0.05 to 0.2 Hz) (J. Healey & Picard, 2000; Lanatà et al., 2015; Munla et al., 2015; Rodrigues et al., 2015).

An increase in a driver's stress level can lead to an increase in LF values and a decrease in HF values. The LF/HF ratio is also considered to be a common stress indicator (Lanatà et al., 2015; Miller & Boyle, 2013; Heikoop et al., 2017). In addition, total power is another feature which is used in the detection of a driver's stress level (Munla et al., 2015).

### Classification

Random Forest (RF) (Breiman, 2001) is an ensemble learning algorithm which provides significant advantages for the classification problem.

An RF classifier is a combination of multitude random forest tree classifiers which each classifier predicts a class label for each input vector. After that, the class label which is selected more than the others will be selected as the final class label. The random trees are grown using different training data sets and random sets of features which increase the diversity of the tree classifiers and create a robust classification model to deal with outliers and noises.

RF has been shown to be an effective classifier for discriminating different stress levels of drivers using physiological signals (Katsis et al., 2008; Soman et al., 2014). Therefore, in this study, the extracted feature set from the HRV signal, using the selected window size and the degree of overlap based on PSO algorithm, is fed to a Random Forest (RF) classifier to classify stress levels. The performance of the proposed model is evaluated using a 10-fold cross-validation.

## IV. EXPERIMENTAL PROCESS

Extensive experiments are conducted to determine if optimising the windowing hyperparameters using the PSO algorithm is an effective strategy to improve the performance of an ECG-based driver stress level detection model. The performance of the proposed framework is evaluated on two different datasets: a public dataset and our dataset. Moreover, the performance of the proposed framework on each dataset is evaluated on different sets of ECG features to investigate the efficacy of different window sizes using different feature sets.

### A. Database

In this study, two datasets are used to evaluate the performance of the proposed framework. The first dataset is a public dataset which contains ECG data collected in real time driving scenarios. The second dataset is collected using an advanced driving simulator in our lab. These datasets are explained in below.

### DRIVEDB dataset

The experiments used a publicly available drivers' stress level detection database collected by Healey and Picard (2000). The database is called "the Stress Recognition in Automobile Driver" database or DRIVEDB and can be downloaded from MIT Media Lab (Healey & Picard, 2008). DRIVEDB contains a collection of different physiological data including ECG, EMG, EDA and RSP from 17 drivers who were driving around for at least one hour in the designed streets and highways in the city of Boston. The sample rate for ECG signal is reported to be 496 Hz. There are three different driving sessions in the dataset, which induced different stress levels. Only 10 drivers were selected for data analysis. It is asked the participants to express their stress levels (low, medium, and high stress) during Rest, highway and City driving.

Table 2 shows the details related to the selected drivers' sessions. Further details of the study protocol can be found in Heikoop et al. (2017). Based on the questionnaire, the least stressful situation is the rest scenario, and the highway driving are more stressful and the city driving is the most stressful situation.

**Table 2.** DRIVEDB dataset information

| Driving session | Stress level | Rating | |
|---|---|---|---|
| | | $\mu$ | $\sigma$ |
| Rest before driving | Low stress | 1.6 | 0.88 |
| Highway | Medium stress | 2.00 | 0.92 |
| City | high stress | 2.55 | 1.02 |

### Our dataset

The dataset used in this study was recorded in response to different stressful situations in the context of driving. The experiment was conducted in an advanced driving simulator (Figure 2). The simulator consists of a car with an automatic transmission, front view (180-degree), rear-view mirrors, audio system, a hydraulic system to simulate vehicle motion and SCANeR™ studio software. The surround-sound such as engine, road noise and other traffic interactions sounds are simulated accurately by the audio system. The physiological signals such as the ECG signal and vehicle dynamics data are acquired using the BIOPAC MP150 and SCANeR™ studio. BIOPAC MP150 was used to acquire physiological signals such as ECG signal with a sampling rate of 1000 Hz. The ECG signal was downsampled to 250 Hz. It should be noted that all the data from the SCANeR™ and BIOPAC software are synchronized. Further information on the of a car with an automatic transmission, front view (180-degree), rear-view mirrors, audio system, a hydraulic system to simulate vehicle motion, and BIOPAC system to acquire ECG signal. Information about the driving simulator can be found in https://research.qut.edu.au/carrsq. In this study we used only the ECG data for analysis.

The data were collected from 27 participants aged 21–40 years (55% male). All participants were required to have a valid Australian driving license, and to regularly drive for a total of at least one hour every week. The total session for each participant took about one hour on average.

### Data collection Scenario

The experimental protocol was structured into two phases: pre-experiment and driving scenario experiment. Prior to the commencement of experiments, all participants were emailed instructed about the experiment (task details, wearable sensors, data acquisition, driving routes, and safety instructions). Some restrictions such as to avoid drinking caffeine and alcohol prior to data collection were applied to the participants.

Before starting driving, each participant was asked to relax for 2–3 minutes to record their physiological baseline. Then, the participant drove through six driving scenarios. Each scenario takes 5 minutes. Along with all the experimental scenarios, the driver's data (physiological, and physical data) as well as the contextual data were continuously acquired. In the first driving scenario (practice drive), the participant was asked to drive on a simple route to become familiar with simulator environment and how to control the car. After the practice drive, the participant drove in the next five driving scenarios: Urban1, Urban2, Highway, City1 and City2 landmarks. It should be noted that the order of the scenarios was randomized across participants to avoid learning effects. Each scenario contains several stressors to induce different stress levels (low, medium, and high) into the participant. The applied stressors in this study were derived from (Hill & Boyle, 2007; Lee et al., 2017; Rodrigues et al., 2015) and were categorized into four groups:

1) Traffic (e.g. driving in heavy traffic)

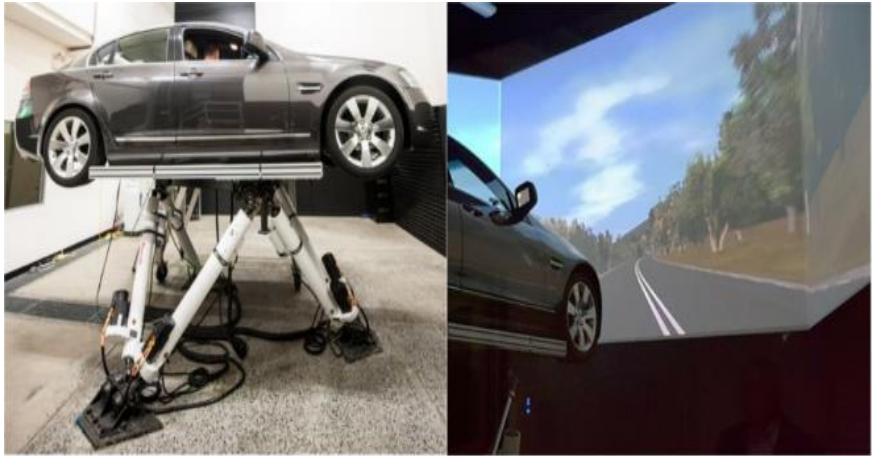

**Figure 2**. CARRS-Q advanced simulator used for our data collection.

2) Difficult driving situations (e.g. narrow roads and tight corners, curved road).

3) Other drivers' behaviors (e.g. angry drivers, careless drivers)

4) Weather and visibility related conditions (e.g. heavy rain, night-time driving, foggy weather).

The intensity and frequency of the stressors used are different in the various scenarios to induce different stress levels. During each scenario, the participants were asked every two minutes to provide their responses (verbally) to a short questionnaire about their average stress levels. They were asked to express their stress levels between 1 and 3 (1-No/low stress to 3- High stress) during each scenario. The Figure 3 presents the distribution of the stress levels for each scenario (Relax, Mountain, Highway, CBD1 and CBD2).

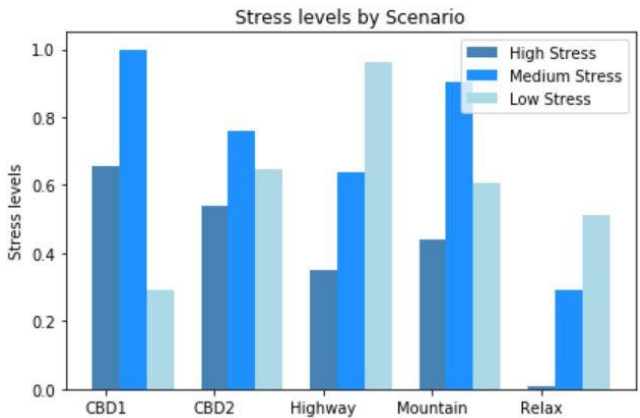

**Figure3.** distribution of stress levels per scenario.

Based on the figure3, the least stressful situation are the relax/rest and highway situations. And the highest stress full situations are CBD1 and CBD2. The Mountain and highway driving are not as stressful as CBD1 and CBD2.

### B. Experimental Setting

Different sets of ECG features were used to evaluate the performance of the proposed framework. The feature sets used in this study are the commonly used ECG features from the time and frequency domains. The feature sets are statistical HRV features, non-linear HRV features and frequency-domain HRV features (See Table 1).

To evaluate the performance of the proposed system based on each ECG feature set, the PSO algorithm was executed 20 times. The PSO algorithm parameters for construction coefficient, damping ratio, particle size are set to 2.05, 0.9, and 5, respectively. These values are taken from our study related to emotion recognition (Nakisa et al., 2017).

In this study, the performance of the proposed system using PSO algorithm was evaluated on two datasets, DRIVEDB and our dataset. The duration of whole experiment for each dataset is different, therefore, the duration of the experiments for each dataset is different. As a result, different window sizes were selected for each dataset. For the DRIVEDB dataset, the window size values were selected to range from 5 to 520 seconds, and the values for the degree of overlap were selected to range from 0 to 95 percent. We tried to select smaller window sizes to build a practical system which can detect driver stress in real-time with high accuracy.

For our dataset, the window size values were selected to range from 5 to 60 seconds, and the values for the degree of overlap were selected to range from 0 to 95 per cent.

It should be mentioned that the number of search iterations was found by trial and error (30 iterations). We have evaluated 100 different values between the ranges of (10,100). The best value which got a good performance regarding to running time and classification accuracy is 30.

The proposed framework in this study was implemented using MATLAB software.

## V. EXPERIMENTAL RESULTS AND DISCUSSION

### A. Evaluating the proposed framework using different ECG feature sets on two our dataset

In this section, the evaluation of the proposed framework based on the three statistical ECG feature sets on DRIVEDB dataset and our dataset is presented.

The performance of the proposed system using PSO was assessed based on the optimum accuracy that can be achieved within an acceptable time. The PSO algorithm was tested based on its ability to find the best windowing hyperparameter values (window size and degree of overlap) within a limited time.

As mentioned in Experimental setting section, the window sizes for the DRIVEDB dataset ranged from 1 to 520 seconds, and for our dataset ranged from 5 to 60 seconds. The reason that the window size of DRIVEDB dataset is different from our dataset is the style of annotation. The annotation of DRIVEDB dataset is done after each experiment (after 520 seconds), while ours is done during the experiment (every 2 minutes). As each experiment/ scenario in our dataset takes 5 minutes, we have considered the annotation for every minute of the experiment. Therefore, the window size for our dataset is between 0 and 60 seconds. As it is mentioned earlier, the performance of the proposed system based on both datasets using different ECG feature set is evaluated over 20 times. The average processing time for each execution is measured by Intel Core i7 CPU, 16 GB RAM, running windows 7 on 64-bit architecture. The average processing time for the proposed method using DRIVEDB dataset and our dataset over 20 runs is around 250 hours. It should be noted that the number of search iterations for each run is 30 iterations and the number of evaluated data points is 150 data points out of 48,925 and 5,225 for DRIVEDB and our collected dataset, respectively. Although the PSO algorithm is computationally expensive, it is less complex than full search algorithms. It should be noted that optimizing hyperparameters to find the best model for driver's stress classification is only conducted during development and training stages.

To study the performance of the PSO algorithm, the window size and the degree of overlap values were segmented into different regions (see Figure 3), and the average accuracy of the solutions based on different regions was calculated. The overlap degree is segmented into three different regions: low (5% -30%), medium (30% -60%) and high (60% - 95%). The average accuracy of solutions for different regions is shown in Figure 4, 5.

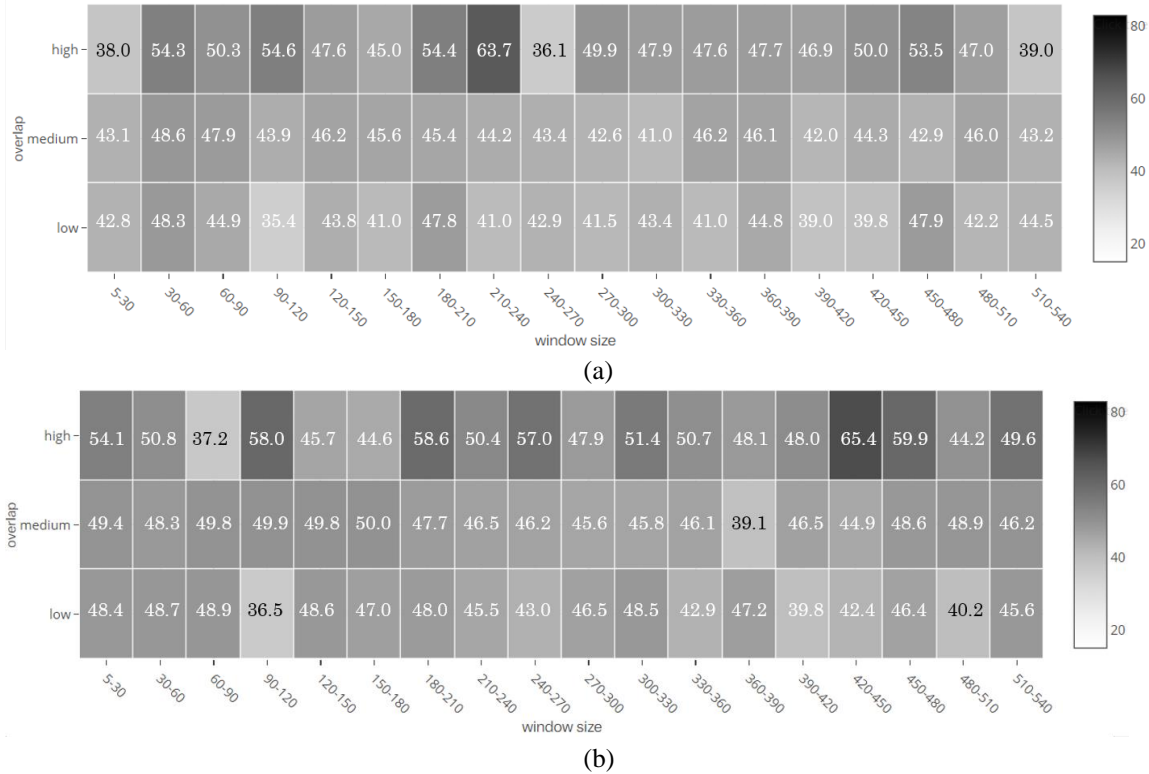

(a)

(b)

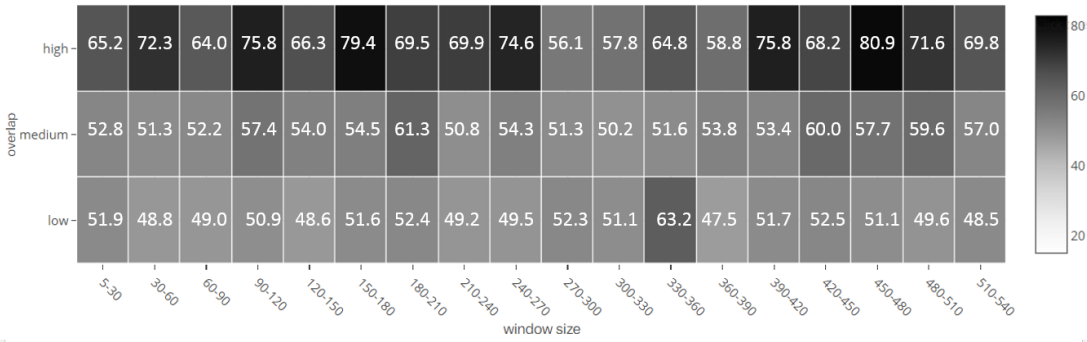

(c)

**Figure 4.** The average performance of the proposed method on DRIVEDB dataset based on different window size and degree of overlap regions using (a) non-linear HRV feature set, (b) frequency-domain HRV feature set, (c) statistical HRV feature set.

Figure 4 presents the average performance of the proposed system based on the DRIVEDB dataset over three different feature sets: non-linear HRV feature set, frequency-domain HRV feature set and statistical HRV feature set. The figure provides the average accuracy for different ranges of degree of overlap and window size.

Based on the average accuracy over three feature sets, the most frequent set of good solutions was found with high degree of overlap (medium to high degree of overlap) and large window size. However, this is not guaranteed, as in some cases a high degree of overlap resulted in a low performance. For example, the minimum average performance using the statistical feature set was obtained with a 90–120 second window and a high degree of overlap. Based on Figure 4 (a), as the size of the window increased, the performance of the proposed method using a non-linear feature set slightly improved. The maximum average performance (63% accuracy) was obtained with a 210–240 second window and a high degree of overlap. It also shows that the accuracy of the proposed method with a medium to high degree of overlap is better than with a low degree of overlap. The minimum performance (35% accuracy) is achieved in a short window size and a low degree of overlap (90–120 second window).

Similarly, the performance of the proposed method based on the frequency feature set increased as the degree of overlap increased (Figure 4 (b)). However, the highest average performance using this feature set was achieved when the window size was long. The best average accuracy was obtained with a 420–450 second window with a high degree of overlap. It should be noted that high degree of overlap has not always resulted in high performance. For example, the minimum average accuracy using the frequency feature set was achieved with a 60–90 second window and a high degree of overlap. This finding shows that the performance of the system depends on the combination of window size and degree of overlap and there are no general values guaranteed to achieve high performance.

As shown in Figure 4 (c), the performance of the proposed hyperparameter search method using the statistical feature set is better than for the non-linear and frequency feature sets and achieved 80% accuracy with a 450–480 second window and a high degree of overlap. It also shows that the performance of driver stress detection using a 150–180 second window and a high degree of overlap is close to the best result.

The performance of the proposed method over the three feature sets was also investigated on our dataset (using advanced simulator) and the results are presented in Figure 5.

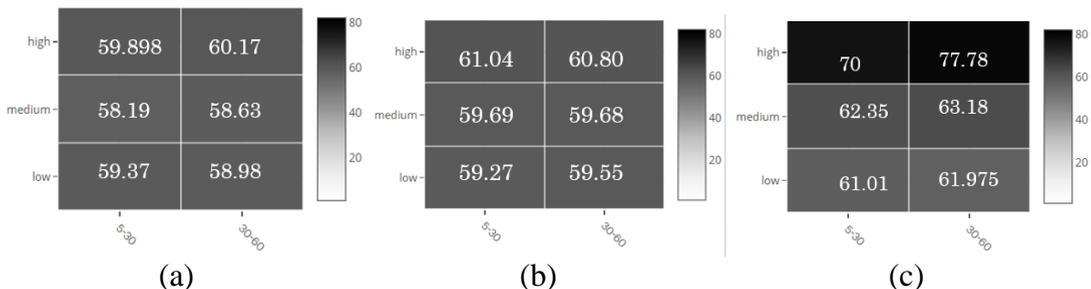

**Figure 5.** The average performance of the proposed method on our dataset based on different window size ranges and degree of overlap ranges using (a) non-linear HRV feature set, (b) frequency-domain HRV feature set, (c) statistical HRV feature set.

**Table 3.** The best performance achieved (maximum accuracy) using the proposed hyperparameter search method based on the three feature sets using the DRIVEDB dataset and our dataset.

| | Statistical feature set | | | Frequency feature set | | | Non-linear feature set | | |
|---|---|---|---|---|---|---|---|---|---|
| | Win size | Overlap | Best Accuracy | Win size | Overlap | Best Accuracy | Win size | Overlap | Best Accuracy |
| DRIVEDB Dataset | 480 | 95 | **92.12%** | 84 | 95 | **68.96%** | 239 | 95 | **70.7%** |
| | 190 | 95 | **89.78%** | 11 | 87 | 50.77% | 30 | 86 | 52.52% |
| | 34 | 95 | 79.50% | 11 | 87 | 50.77% | 11 | 54 | 47.90% |
| | 10 | 95 | 71.66% | 5 | 50 | 50.02% | 10 | 50 | 46.88% |
| Our dataset | 60 | 95 | **77.78%** | 45 | 95 | **61.34%** | 5 | 95 | **61.97%** |
| | 30 | 95 | **74.63%** | 5 | 95 | 60.91% | 39 | 16 | 61.94% |
| | 10 | 95 | 68.86% | 60 | 10 | 60.4% | 45 | 95 | 61.17% |
| | 5 | 95 | 66.51% | 45 | 10 | 59.36% | 47 | 76 | 60.16% |

The result using a non-linear feature set (Figure 5 (a)) shows that the best average performance is obtained with a 30–60 second window and a high degree of overlap. The performance of the stress detection using our dataset is 9% better, compared to the DRIVEDB dataset that achieved 51% accuracy with a 30–60 second window. Based on the results, the performance of the proposed method using the frequency feature set is better than using the non-linear feature set and it achieved 61% accuracy over short window sizes (5–30 seconds). Similar to the DRIVEDB dataset, the highest average performance is obtained using the time-domain feature set with a 30–60 second window and a high degree of overlap. Using our dataset, we achieved 77.7% accuracy, while using the DRIVEDB dataset we achieved 80% accuracy.

Further analysis of the best solutions achieved using the proposed method is provided in Table 3. The table presents the highest accuracies achieved using three feature sets over DRIVEDB and our dataset.

As shown in Table 3, the highest performance based on the statistical feature set using the DRIVEDB dataset is 92% with a long window size and a high degree of overlap (480-second window and 95% overlap). It also shows that all the high performances are achieved with a high degree of overlap. Similarly, the best accuracies over our dataset are obtained with a high degree of overlap. Based on both the DRIVEDB dataset and our dataset using the statistical feature set, the results show that longer window sizes resulted in higher accuracies.

The best results based on the frequency feature set have short window sizes. Using the DRIVEDB dataset, the best accuracy is achieved with an 84-second window and the best window size ranges from 11 to 80 seconds.

The best window sizes over using dataset range from 5 to 65 seconds and the highest accuracy is achieved with shorter window sizes (45 seconds).The best accuracies achieved for the non-linear feature set are lower than for the frequency and statistical feature sets. The results show that the best accuracy achieved using the DRIVEDB dataset is 70.7% with large window sizes and a high degree of overlap. Moreover, as the size of windows decreased, the performance of the proposed method using non-linear features decreased. In contrast to the DRIVEDB dataset, the best accuracies using our dataset were obtained with short window sizes. Moreover, the performance of the proposed method using our dataset is close to the performance using the DRIVEDB dataset.

**Table 4.** The best performance achieved (maximum accuracy) using the Random Search and PSO algorithm based on the three feature sets using the DRIVEDB dataset and our dataset.

| Hyperparameter optimization | | Random Search | | | PSO algorithm | | |
|---|---|---|---|---|---|---|---|
| | | Win Size | Overlap | Accuracy (Best) | Win Size | Overlap | Accuracy (Best) |
| DRIVE DB Dataset | Statistical feature set | 230 | 90 | 83% | 480 | 95 | **92.12%** |
| | Frequency feature set | 120 | 90 | 59.46% | 84 | 95 | **68.96%** |
| | Non-linear feature set | 90 | 85 | 55.84% | 239 | 95 | **70.7%** |
| Our Dataset | Statistical feature set | `55 | 87 | 65.96 % | 60 | 95 | **77.78%** |
| | Frequency feature set | 49 | 95 | **61.46%** | 45 | 95 | 61.34% |
| | Non-linear feature set | 45 | 50 | 45.47% | 5 | 95 | **61.97%** |

## B. Evaluating windowing hyperparameter optimization using PSO and Random Search algorithms

The performance of the proposed system using PSO algorithm is evaluated and compared with Random Search algorithm in this section. The models are built and compared on two datasets, DRIVEDB and our dataset in this section. To apply Random Search algorithm, Hyperopt library, Random.suggest algorithm is used. The other settings are discussed in the experimental setting section.

Table 4 presents the performance of the proposed model using PSO and Random search model, providing the optimum accuracy that can be achieved within an acceptable time. Based on the achieved results from the previous section, the performance of Random Search and PSO model are evaluated based on statistical features. Processing time is determined by Intel core i7 CPU, 16GB RAM, running windows 7 on 64bit architecture. Based on the best obtained accuracy, the general performance of Random Search algorithm was slightly lower than PSO algorithms.

The best accuracy using both Random Search and PSO algorithm are achieved with large window size and high degree of overlap. Moreover, statistical features using both Random Search and PSO algorithm is performing better than other feature sets (Frequency and Non-linear feature sets).

Based on the results from Table 3 and 4 we can conclude that PSO algorithm is more successful in finding the near optimal windowing hyperparameters. This is because PSO algorithm is capable in maintaining high diversity in exploring and finding better solutions.

Random search algorithm is slightly less distributed in the whole search space and for more evenly distributed in subspaces.

## C. Comparison of the best model built by the framework with other recent works

In this section, we compare the performance of the best built model against existing studies in the literature (see Table 5). The studies reported in Table 5 used ECG modality to build an accurate model that can detect driver stress levels using the DRIVEDB dataset and they are compared with our proposed method. In this table we compared different studies based on the highest classification accuracy for different numbers of stress levels as well as different classifiers.

Based on the Table 5, Wang et.al (2013) and Keshan et al. (2015) achieved promising results (about 97%) using ECG signal; however, these methods were used for a two-level classification (low and high stress). while our method was able to achieve 92% accuracy over a three-level classification (low, medium, and high). The performance of their methods was tested on only the DRIVEDB dataset, whereas our proposed methods were tested on the DRIVEDB dataset as well as our dataset. Wang et al. (2013) classified three stress levels using a MLP classifier with large window sizes, between 840 and 2280 seconds long, while our methods achieved higher accuracy (92%) with shorter window sizes (480 seconds). Compared to all the latest studies, our proposed methods produced the state-of-the-art performance over both DRIVEDB and our dataset, which confirms the value of windowing hyperparameter optimization to improve stress classification. The results confirm that optimizing both windowing hyperparameters (window size and the degree of overlap) are essential to achieve an accurate driver stress detection model.

## D. Discussion

The results of this study indicate that optimising the window size and degree of overlap hyperparameters is an effective strategy to build an accurate ECG-based model.

This hyperparameter shows that as the degree of overlap increased the time variation of ECG signal can be captured finer and improved the performance of classification. Based on the presented results, previous studies that used physiological signals in heath domains have demonstrated that increasing the degree of window overlap can improve the detection performance (Delachaux et al., 2013; Janidarmian et al., 2014; Fekr et al., 2016).

Our study also confirms that statistical HRV features are reliable indicators of stress levels. This finding supports previous research (Lee et al., 2007), which suggested that statistical HRV time-domain features are good indicators of instantaneous driver stress responses. Another important finding is that the most accurate models, for each ECG feature set, are built based on large window sizes. This means that the extracted features for long-term HRV variations can better reflect dynamic ANS activities related to stress, which is in line with previous results (Wang et al., 2013; Keshan et al., 2015; Bichindaritz et al., 2017).

Table 5. Comparison of our approach with other latest works.

| Reference | Experiment setting | Physiological Signal used | Classifier | Windowing Hyperparameter | | Performance (Accuracy) | No. Classes |
| --- | --- | --- | --- | --- | --- | --- | --- |
| | | | | Window size | Overlap | | |
| Wang et al., 2008 | DRIVEDB Dataset (time and frequency from HRV) | ECG | SVM | 300 seconds | 50% | 97.5% | 2 stress classes (low, high) |
| Keshan et al. 2015 | DRIVEDB Dataset (statistical from RR intervals) | ECG | Decision tree (J48) | 840-2280 seconds | - | 97.92% | 2 stress classes (low, high) |
| | | | | | | 68.66% | 3 classes (low, medium, and high) |
| Munla N., 2015 | DRIVEDB Dataset (time domain feature) | ECG | SVM | 300 seconds | - | 83.33% | 2 Stress classes (stress/ no stress) |
| Wang et al. 2013 | DRIVEDB Dataset (statistical and non-linear from RR intervals) | ECG | MLP | 840-2280 second | - | 80% | 3 stress classes (low, medium, and high) |
| Our work | DRIVEDB Dataset (statistical HRV feature set) | ECG | RF | 480 seconds | 90% | 87% | 3 stress classes (low, medium, and high) |
| | Our Dataset (statistical HRV feature set) | ECG | RF | 60 seconds | 90% | 77.78% | |

## VI. CONCLUSION

In this study we improved the performance of driver stress detection model by proposing an efficient framework to optimise windowing hyperparameters (window size and degree of overlap). To optimise these hyperparameters and build an accurate driver stress detection model, we adapted the PSO algorithm. The performance of the proposed framework was evaluated to detect three stress levels of drivers (low-level, medium-level, and high-level) using different ECG feature sets. The proposed feature sets contained the common extracted features related to driver stress level from HRV signals from the time and frequency domains. Two datasets (DRIVEDB and our dataset) are used to evaluate the performance of the proposed hyperparameter search model. DRIVEDB dataset is a public dataset that was collected from real time driving scenarios by MIT Media Lab and our dataset was collected from an advanced driving simulator.

We conducted comprehensive results showing that optimising the window size and the degree of overlap hyperparameters is a key step in the process of building an accurate driver stress level detection model. The proposed framework was successful in building a driver stress level detection model and achieved the state-of-the-art performance with 92.12% and 77.78% accuracies over DRIVEDB dataset and our dataset, respectively.

This study confirms that ECG signal can be used to detect driver stress level with high accuracy. In our view, the obtained results are an excellent initial step towards building a practical system using a single physiological signal that can continuously and automatically detect driver stress level. The findings in this study need to be generalised on more datasets. In addition, it would be interesting to investigate the efficiency of other physiological signals to build such a system using the proposed framework.

### ACKNOWLEDGMENT

Frederic Maire acknowledges continued support from the Queensland University of Technology (QUT) through the Centre for Robotics.